\documentclass[12pt,draft]{article}
\usepackage{amssymb}
\usepackage{latexsym}
\input psfig.sty

\newcommand{\N}{{\mathbb N}}
\newcommand{\R}{{\mathbb R}}
\newcommand{\dd}{{\rm d}}

\font\fone=cmr10 scaled\magstep3
\font\ftwo=cmr7 scaled\magstep3
\begin{document}
\thispagestyle{empty}
\vspace{1.5in}

\centerline{\fone Non-Turing computations via Malament--Hogarth
space-times}
\vspace{0.3in}

\centerline{\ftwo G\'abor Etesi\footnote{Research Supported by the Japan
Society for Promotion of Science, grant No. \ P99736.}}
\vspace{0.1in}

\centerline{Yukawa Institute for Theoretical Physics,}
\centerline{Kyoto University,}
\centerline{606-8502 Kyoto, Japan}
\centerline{\tt etesi@yukawa.kyoto-u.ac.jp} 
\vspace{0.2 in}

\centerline{\ftwo Istv\'an N\'emeti\footnote{Research supported by 
Hungarian National Foundation for Scientific Research grant No.\ T30314.}}
\vspace{0.1in}
\centerline{Alfr\'ed R\'enyi Mathematical Institute of the }
\centerline{Hungarian Academy of Science}
\centerline{Re\'altanoda u. 13-15, Budapest}
\centerline{H-1053 Hungary}
\centerline{{\tt nemeti@renyi.hu}}
\vspace{0.2 in}

\begin{abstract}
We investigate the Church--Kalm\'ar--Kreisel--Turing Theses concerning
theoretical (necessary) limitations of future computers and of deductive
sciences, in view of recent results of classical general relativity theory. 
We argue that (i) there are several distinguished Church--Turing-type 
Theses (not only one) and (ii) validity of some of these theses depend on
the background physical theory we choose to use. In particular, if we
choose classical general relativity theory as our background theory, then
the above mentioned limitations (predicted by these Theses) become no more
necessary, hence certain forms of the Church--Turing Thesis cease to be
valid (in general relativity). (For other choices of the background theory
the answer might be different.) 

We also look at various ``obstacles'' to computing a
non-recursive function (by relying on relativistic phenomena) published in
the literature and show that they can be avoided (by improving the
``design'' of our future computer). We also ask ourselves, how all this
reflects on the arithmetical hierarchy and the analytical hierarchy of
uncomputable functions.
\end{abstract} 
\vspace{0.1in}

\centerline{Keywords: {\it Church--Turing Thesis, Computers, Black Holes}}
\pagestyle{myheadings}
\markright{G. Etesi, I. N\'emeti: Black Holes and the Church--Turing
Thesis}
\section{Introduction}
Certain variants of the so-called Church--Turing Thesis play a basic role
in the foundations of theoretical computer science, logic,
meta-mathematics and the so-called ``fundamentals of deductive
sciences''. The Thesis is a well-reasoned, well-motivated
``conjecture''
\footnote{We mean the kind of conjecture which cannot be
proved but can, in principle, be refuted.}. 
The Thesis was formulated before
``black hole physics'' was developed. We will recall the Thesis and
some of its variants in detail in Section 2.

Roughly speaking, the variant we are interested in concerns inherent
limitations of possible future computing devices. These limitations deal
with {\it idealized} computers, therefore they do not involve particular
data like the size of our Universe, etc. i.e. these limitations are
supposed to be necessarily (i.e. theoretically) true in some sense. 
On the other hand they do involve some physical theory about time, space,
motion and things like that as we will argue in Section 2. Clearly, if we
do not presuppose a consistent theory about time, space, motion etc. then
it is impossible to formulate Theses of the kind we are referring to. Very
roughly, these variants of the Church--Turing Thesis conjecture that if a
mathematical function $f$  will be realizable at least in principle by an
arbitrary future ``artificial computing system'' then $f$ must be
Turing-computable\footnote{This is only a first, incomplete approximation 
of a part of the Church--Kalm\'ar--Turing Theses, however. We refer to
Theses 2-4' in Section 2 for
a better illustration of what the theses we want to discuss here are
about.}. Here again, the future ``artificial computing system'' is
understood as being {\it idealized} and $f$ is realized by the system if
{\it in theory} it is realized by the theoretical description (design) of
the system. 

In passing we note that this Thesis has many important consequences. One
consequence says that ``paper-and-pencil-computability'' coincides
(and will always coincide) with machine-computability. Here by
paper-and-pencil-computability we understand realizability by an {\it
algorithm} in the mathematical sense (where we note that the mathematical
notion of the algorithm goes back to ancient Greeks, in some sense).

{\it Remark.} Some authors, e.g. Pitowsky \cite{pit} argue that the Thesis
we are interested in is not really Church's Thesis but Wolfram's Thesis
(cf. [Wolfram 1985] from the references in \cite{pit}). The argument
states that Church was not interested in computers, but instead he was
interested in the ``purely mathematical'' notion of an algorithm. We 
would like to pose the following counter-arguments to 
this objection collected in items (1)-(5) below.

(1) 
It is exactly this subtlety because of which we refer to those
variants of the Thesis we want to discuss here as {\it
Church--Kalm\'ar--Turing Theses}\footnote{Actually we should call them as
Church--Kalm\'ar--Kreisler--Turing Theses but for simplicity we will
write Church--Kalm\'ar--Turing Theses.} (instead of calling it Church's
or Church--Turing Theses).
Perhaps Church himself was not interested in computers but Kalm\'ar and
Turing were, and they did take part (emphatically) in refining, 
publishing etc. of the ``abstract, idealized, theoretical, 
future-computer-oriented'' version of the Thesis\footnote{One of the
present authors (I.N.) knew Kalm\'ar in person, and he remembers that it
was the present variants (cf. Theses 2-4' below) of the Church--Turing
Thesis Kalm\'ar was most interested in.}.  

(2) Independently of Church's original motivation, if we look into the
literature of our natural sciences today, we find that in the branches
listed in the beginning of this introduction (e.g. theoretical computer
science, artificial intelligence, cognitive science, etc.) the ``abstract
computer-oriented'' version of the Thesis is being used essentially under
the name ``Church's Thesis'', cf. Odifreddi \cite{odi}, I.8. (pp.
101-122), Gandy \cite{gan}, Kreisel \cite{kre}. Consequently we think that
it is completely justified to investigate the question of under what
assumptions these ``incarnations'' of the Thesis are valid and it is
reasonable then to refer to these incarnations as (variants of) Church's
Thesis.

(3) We quote from the textbook of Odifreddi \cite{odi}, p. 51: ``Turing
machines are theoretical devices, but have been {\it designed with an
eye on physical limitations}''. Hence, if we are talking  about the
Church--Turing Thesis (as is quite customary) then we cannot
agree with Pitowsky's and others claim that the Thesis would be only about
the purely mathematical notion of algorithms and would have nothing to do
with the limitations of idealized physical computers. (Actually, Gandy
\cite{gan} investigated in some detail the ``idealized physical computer''
aspect of the Church's Thesis.)

(4) The issue whether the Church--Turing Thesis is only about the pure
mathematical notion of an algorithm or whether it also concerns the
theoretical limitations of idealized, abstract computing devices (based on
some physical theory), has been discussed extensively in the literature of
theoretical computer science, logic and related fields. E.g. a special
issue of the Notre Dame Journal of Formal Logic \cite{not} is devoted to
the subject. We cannot quote all the relevant references  here but many of
them can be found in \cite{odi}, I.8, scattered over pp. 101-123.

The general conclusion is that the Church--Turing Thesis is not one Thesis
but a collection of {\it several} Theses\footnote{cf. e.g. \cite{odi} p.
123.} some of which deal with the purely mathematical concept of
algorithms while other (just as respectable) ones concern (among others)
the theoretical limitations of idealized future computing systems which
subject will be further elaborated in a more unambiguous manner in Section
2 below.

(5) In \cite{odi}, p. 103 one can read that in meta-mathematics Church's
Thesis is used to prove ``absolute unsolvability''. To our minds this
clearly points in the direction we want to go; namely if a problem is
decidable by performing a thought-experiment (consistent, say, with the
classical general relativity) then the problem is not absolutely
unsolvable (nevertheless it may remain unsolvable for various reasons like
lack of resources). We finished our remarks concerning Pitowsky's
objection.

The notion of ``computable function'' splits up into at least three
notions. These are 

(i) computability by a pure mathematical algorithm (in 
the purely mathematical sense); 

(ii) computability by some idealized, future computing device based on
{\it some} physical theory (like classical general relativity or quantum 
mechanics, etc.);

(iii) computability by some computing device based 
on our present physical world-view i.e. taking into
account {\it all} of our present day physical, cosmological, etc.
knowledge on the Universe we are living in. 
   
We would like to illustrate that distinctions between
(i)-(iii) are reasonable and not trivial by the following.

In connection with the distinction between (i) and (ii) we note that 
if we want to define paper-and-pencil-computability done by a group of
mathematicians, the question comes up whether we allow one of the
mathematicians to take an air trip during the course of their computations
(or take a trip by a space-ship to a rotating black hole); if we say yes 
we need to select a physical theory to control these motions.

Our intended, main distinction between (ii) and (iii) above is that 
in (ii) physical theories are considered as sets of consistent physical
laws without initial data in contrast with (iii) where particular initial
data are also taken into account (and the most general known physical
theory is used). Furthermore we emphasize that ``selection of a particular
physical theory'' in (ii) is acceptable from science-historical viewpoint
only, i.e. without taking into account the particular development of
physical sciences we have no reason to choose a certain physical theory,
we always should use the whole present physical world-view. Note also that
by lack of ``monotonity'' of the development of physical theories, by
selecting a certain theory, we have to
face the fact that our statements within the framework of the chosen
theory may not continue to hold in a more general (future)
theory\footnote{E.g. in {\it classical electrodynamics} one deduces that
electrons must emit electromagnetic radiation while orbiting around
nuclei; this statement is not true in a more general theory, called 
{\it quantum mechanics}.}. 

We will call the ways of computability listed in (i)-(iii) as
{\it computability of the first kind}, etc., respectively. In the present
work we want to show (among other things) that computability of the first
kind and second kind are {\it not} necessarily equivalent\footnote{In
principle this non-equivalence could be attacked by the approach of the
school of Pour-El et al \cite{pou-ric} but here we are ``more ambitious''
in the sense that we want to keep our computers ``programmable and
logic oriented'' (i.e. ``digital'' as opposed to ``analog'')
explained more clearly in Section 3. We will show the non-equivalence
by describing idealized future computing devices (e.g. in Proposition
1) which realize functions not Turing computable. We think
computability of the first kind cannot be too different from Turing
computability (and our second kind computable functions in Section 3
are rather far from being Turing computable).}.

Further, we note that computability of the third kind does not fit
smoothly with present day computability theory in the sense that most
Turing-computable functions are not computable of the third kind (e.g. by
lack of enough time for a huge calculation if the Universe has finitely
long future only). Hence in the present work we do not want to discuss
computability of the third kind, while we acknowledge that it is a
potentially interesting subject. We note that in our opinion the most
emphatically used obstacle in \cite{pit} applies only to computability of
the third kind, hence it does not apply to the main subject of this paper
which is computability of the second kind. (We also note that the famous
classical theorists of the field e.g. Kalm\'ar, Kreisler, Turing, were more
interested in computability of the second kind than in the third kind, 
in our opinion).

Our paper is organized as follows. In Section 2 we will recall and discuss
the above mentioned variants or incarnations of the
Church--Turing Thesis (called Church--Kalm\'ar--Turing Theses). Then, in
Section 3, we will raise the question whether within the framework of
classical general relativity theory some forms of the
Church--Kalm\'ar--Turing Theses admit a
counterexample. We will find that, very probably, such a counterexample is
possible, at least in theory. Both in \cite{ear} and in \cite{pit} there
are some obstacles to the possibility of such kinds of counterexamples. We
will look at these obstacles one by one in Section 4 and will argue that
they can be avoided in the case of a certain thought-experiment (i.e. a
certain ``design of the idealized future computing device''). E.g. we will
argue that the observer who will find out the solution of an 
``unsolvable problem''\footnote{For instance the consistency of ZFC set
theory can be such a problem.} does not have to pay with his destruction
for accessing this piece of knowledge.

The basic ideas elaborated in this paper have been around for a while. For
example in the academic year 1987/88 at the University of Iowa in Ames
(USA) one of the present authors gave a course in which these ideas
were discussed\footnote{Iowa State University, Department of Mathematics.
Ph. D. course during the academic year 1987/88. Subject: "On logic,
relativity, and the limitations of human knowledge."
Lecturer: I. N\'emeti.}, see also \cite{and1} \cite{and2}; in 1990
Pitowsky in \cite{pit} considered such
ideas in a slightly more pessimistic spirit, and in 1995 Earman in
\cite{ear} examined such ideas under the name of constructibility or
possibility of Plato machines (\cite{ear} pp. 101-123). However, the
emphasis in Earman's book and other works like \cite{ear-nor1},
\cite{ear-nor2} and \cite{ear-nor3} is more on ``supertasks'' rather than
on the Church--Kalm\'ar--Turing Theses. Other related work 
we mention is Gr\"unbaum's \cite{gru} while a quantum mechanical idea is
worked out in \cite{kie} related with Hilbert's tenth problem. This list
of references is far from being complete, e.g. we should have mentioned
the important papers of Hogarth \cite{hog1}\cite{hog2} which will be
essential in our considerations. Hamkins and Lewis \cite{hale} is a recent
paper in this direction. 

In view of the above, the purposes of the present paper are the
following: (i) put the emphasis on the Church--Kalm\'ar--Turing Theses
(instead ofe.g. supertasks) in a thorough, systematic way; (ii) formulate exactly
which versions of the Church--Kalm\'ar--Turing Theses 
we want to investigate (and what do they mean); (iii) formulate carefully
what we understand under a counterexample for these variants; (iv) see if
the apparent obstacles e.g. listed in earlier works can be avoided (at
least in theory).

\section{The Church--Kalm\'ar--Turing Theses}
In this section we formulate some variants of the Church--Turing
Thesis based on the hierarchy of definable functions $f:\N\rightarrow\N$.
We follow notation and definitions of \cite{odi}. Thesis 1 below is only
the first approximation of the Church--Kalm\'ar--Turing Theses we want to
investigate; therefore beyond Thesis 1 we will use more unambiguous, more
carefully specified, more tangible formulations-variants of the Theses.
These will be Thesis 2-2' and 3 (Theses 4-4' are for completeness only).

Let $X$ be a finite set and denote by $X^*$ the set of finite sequences
over $X$. For sake of convenience we choose $X:=\{ 0, 1\}$. 
\vspace{0.1in}

{\bf Definition 1.} We call a function $f: X^*\rightarrow X^*$ {\it
Turing-computable} if there is a Turing machine which realizes $f$.
$\Diamond$
\vspace{0.1in}

\noindent For the definition of a Turing machine, see Definition I.4.1
while the realization of a function by a Turing machine is formulated in
Definition I.4.2 of \cite{odi}. 

As it is well-known, the set of natural numbers, 
$\N=\{ 0,1,2,\dots\}$, can be represented
as $X^*$ i.e. there is a bijection $\N\cong X^*$ which is effectively
computable in the intuitive sense. Consequently the notion
of a Turing-computable number-theoretic function $f:\N\rightarrow\N$  is
well-defined i.e. Turing-computability of these functions is independent
of the representation of $\N$ as an $X^*$
\footnote{We could introduce the
notion of a {\it recursive function} $f: \N\rightarrow\N$
as well (see the various definitions in Chapter I of \cite{odi}). But
according to a theorem of Turing (e.g. Theorem I.4.3 of \cite{odi}) a
function $f: \N\rightarrow\N$ {\it is Turing-computable if and only if
it is recursive}, hence we will use the term ``Turing-computable'' 
systematically throughout this paper.}. 

Introducing the notation
\[\N^k:=\underbrace{\N\times\dots\times\N}_k\:\:\:\:\:\mbox{ for }k\in\N^+
,\]
where $\N^+=\{ 1,2,\dots\}$ denotes the set of positive integers, 
we can see that $\N^k$ can also be regarded as a subset of $Y^*$ where $Y$
contains some extra element in comparison with $X$, for example $Y:=\{
0,1,-\}=X\cup\{ -\}$. As an example, $101-11\in Y^*$ corresponds to the
pair $(5, 3)\in\N^2$ in this notation. In this way we can talk about the
Turing-computability of a function $f :\N^k\rightarrow\N^m$ for each
$k,m\in\N^+$. 
\vspace{0.1in}

{\bf Definition 2.} A subset $R\subseteq\N^m$ is called an {\it ($m$-ary)
relation}.

(i) A relation $R\subseteq\N^m$ is called {\it decidable} if its
characteristic function $\chi_R: \N^m\rightarrow\{ 0,1\}$, given by
\[\chi_R(x_1,\dots ,x_m):=\left\{ \begin{array}{ll}
                               1 & \mbox{if $(x_1,\dots ,x_m)\in R$}\\
                               0 & \mbox{if $(x_1,\dots ,x_m)\notin R$,}
                             \end{array}
                     \right. \]
is Turing-computable;
   
(ii) A relation $R\subseteq\N^m$ is called {\it recursively enumerable} if
there is a Turing-computable function $f_R: \N\rightarrow\N^m$ such that
${\rm im}f_R=R$ where in general
\[{\rm im}\:f:=\{ (y_1,\dots ,y_m)\:\vert\:\exists\:
x\:\:f(x)=(y_1,\dots ,y_m)\} \]
is the {\it image} of a function $f:\N\rightarrow\N^m$. $\Diamond$
\vspace{0.1in}

\noindent In this way we have defined decidable and recursively enumerable
$m$-ary relations for all $m\in\N^+$. Next we introduce a natural hierarchy
from the computability viewpoint on the set of relations.
\vspace{0.1in}

{\bf Definition 3.} Let $R\subseteq\N^m$ be a relation. 

(i) We say that the relation $R\subseteq\N^m$ is
a {\it $\Sigma_1$-relation} i.e. $R\in\Sigma_1$ if $R$ is recursively
enumerable;

(ii) We say that the relation $R\subseteq\N^m$ is a {\it
$\Pi_1$-relation} i.e. $R\in\Pi_1$ if $\overline{R}\in\Sigma_1$. Here
$\overline{R}:=\N^m\setminus R$ is the complement of $R$ with respect to
$\N^m$;

(iii) In general, we say that a relation $R\subseteq\N^m$ {\it is a
$\Sigma_n$-relation} i.e. $R\in\Sigma_n$ ($n\in\N, n\ge 2$) if there is a $k\in\N$
and a $\Pi_{n-1}$-relation $S\subset\N^{m+k}$ such that 
\[ R=\{ (x_1,\dots ,x_m)\vert\:\exists\:(x_{m+1},\dots ,x_{m+k})\in\N^k,\:
(x_1,\dots ,x_{m+k})\in S\} .\] 

(iv) In general, we say that a relation $R\subseteq\N^m$ {\it is a
$\Pi_n$-relation} i.e. $R\in\Pi_n$ if $\overline{R}\in\Sigma_n$.
$\Diamond$ 
\vspace{0.1in}

\noindent We will use $\Sigma_n$ also as the set of all
$\Sigma_n$-relations,
and similarly for $\Pi_n$. Thus
e.g. $R\in\Sigma_2\setminus(\Sigma_1\cup\Pi_1)$ means that
$R\in\Sigma_2$ but $R\notin\Sigma_1$ and $R\notin\Pi_1$,
i.e. $R\in\Sigma_2$ and neither $R$ nor its complement is recursively
enumerable.

Notice that every function $f: \N^k\rightarrow\N^m$ may be
considered as a relation
\[ R_f:=\{ (x_1,\dots ,x_k, y_1,\dots ,y_m)\:\vert\:
f(x_1,\dots ,x_k)=(y_1,\dots ,y_m)\}\subset\N^{m+k}.\]
$R_f$ is called the {\it graph} of $f$. We will say that a function
$f:\N^k\rightarrow\N^m$ is a {\it $\Sigma_n$-function} (resp. {\it
$\Pi_n$-function}) if and only if its graph $R_f$ is a $\Sigma_n$- (resp.
$\Pi_n$-) relation.

By keeping in mind the
definition of Turing machines, one can easily show the following (see
e.g. \cite{odi}): 

\begin{enumerate}
\item[(i)]
A function $f$ is Turing-computable if and only if its graph $R_f$ is
recursively enumerable, i.e. if $R_f\in\Sigma_1$.
\item[(ii)]
A relation $R$ is decidable if and only if both $R$ and its complement
are recursively enumerable, i.e. if and only if
$R\in\Sigma_1\cap\Pi_1$.
\end{enumerate}
Thus, $R\in\Sigma_1\setminus\Pi_1$ means that $R$ is recursively
enumerable but $R$ is not decidable. As an example, one may consider the
relation $D_e$ defined by a Diophantine equation $e(x,y,a)$ as follows
\[D_e:=\{ (x,y)\:\vert\:\exists\:a\:\:e(x,y,a)\}\subset\N^2 \]
which is clearly $\Sigma_1$ but not necessarily $\Pi_1$ i.e. it is not
necessarily decidable although recursively enumerable. Indeed, there are
choices of  the equation $e(x,y,a)$  for which 
$D_e$ is undecidable. 
One can see that there are relations in $\Sigma_2$ which are not
recursively enumerable because of using existential quantifications in
their definitions. In general, there are $\Sigma_n$-relations which
are not $\Sigma_{n-1}$-relations, and, intuitively, the
$\Sigma_n$-relations are ``harder to compute'' than the
$\Sigma_{n-1}$-relations. The sets $\Sigma_n$, $\Pi_n$ measure
the degree of {\it non-computability} of a relation by means of Turing
machines i.e. algorithms. For details see Chapter IV of \cite{odi}.
 
The hierarchy $\Sigma_1 ,\Pi_1 ,\dots ,\Sigma_n ,\Pi_n,\dots$
$(n\in\N^+ )$ is called {\it arithmetical hierarchy}. It provides us
subsets $R\subset\N^m$ which are further and further away from being
computable. Beyond the arithmetical hierarchy comes the so-called {\it
analytical hierarchy}. We note that the first order logic theory
$Th(\langle\N,+,*\rangle )$ of arithmetics is at the bottom of the
analytical hierarchy.

At this point one may raise the question if there is a hypothetical
{\it extended Turing machine} such that all the elements of $\Sigma_1$
would become decidable by this machine. Such an extended Turing machine
should possess only one extra property compared to the ordinary Turing
machines. Indeed, it should be able to answer the following
question in finite time: does a given ordinary Turing machine 
stop with a given input $y$ or not? Such an extended Turing machine
certainly exists as an abstract mathematical object but it may or may not
be realized physically.

It is possible to show that, by using this one extra
ability, all elements of $\Sigma_1$ become decidable
(in the extended sense) while elements of $\Sigma_2$ would become recursively
enumerable (in the extended sense). This means that, by using these
extended Turing machines, every relation would become ``less
non-computable with one unit''.
  
The concept of a Turing machine is an extraction, idealization or an 
abstract formulation of our experience with physical computers. By a {\it
physical computer (in the narrow sense)} 
we mean a discrete physical system together with a
physical theory for its behaviour (see \cite{odi}, p. 104). Hence one may
ask if the above mentioned extended Turing machine can be realized as a
physical computer. We will say that a function $f:\N^k\rightarrow\N^m$ is
{\it effectively computable} if there is a physical computer realizing it.
Here by ``realization by a physical computer''we mean the following.

Let $P$ be a physical computer, and $f: \N^k\rightarrow\N^m$  a
(mathematical) function. Then we say that  $P$  {\it realizes}  $f$  if an
imaginary observer
$O$  can do the following with  $P$. Assume  $O$ receives an 
arbitrary element
$(x_1,\dots ,x_k)\in \N^k$ from, say, his ``opponent''. Then  $O$  can
``start'' the computer  $P$  with $(x_1,\dots ,x_k)$  as an input and then
sometime later (according to $O$'s internal clock) $O$ ``receives'' data
$(y_1,\dots ,y_m)\in\N^m$  from  $P$  as an output such that
$(y_1,\dots ,y_m)$ coincides with the value  $f(x_1,\dots ,x_k)$  of the
function $f$ at input $(x_1,\dots ,x_k)$. The reason why we wrote
``start'' and ``receives'' in quotation marks is that we do not want to
specify how $O$  can start  $P$  etc., these can be specified by the
designer of the computer $P$.  The essential idea is that  $O$  can use
$P$  as a device for computing $f$. The difficulty which we have to
circumnavigate (when defining what we mean by saying ``$P$ realizes $f$'') 
is that $f$ is an infinite object. The solution is that we postulate that
for {\it any} permitted choice of the input data $(x_1,\dots ,x_k)$,
computer
$P$ will produce an output $(y_1,\dots ,y_m)$ 
and in addition, this output will coincide with
$f(x_1,\dots , x_k)$. We emphasize that this definition does not require
repeated activations of  $P$ , instead it says that whatever inputvalue
$(x_1,\dots ,x_k)$ we would choose,  $P$  will produce an output
coinciding with  $f(x_1,\dots , x_k)$.

In this context we may quote the original form of the Church--Turing
Thesis (\cite{odi} p. 102):
\vspace{0.1in}

{\bf Thesis 1} (Church--Turing). {\it Every effectively computable
function $f:\N^k\rightarrow\N^m$ gives rise to a relation $R_f\in\Sigma_1$
i.e. every effectively computable function is Turing-computable.}  
\vspace{0.1in}

\noindent In light of Thesis 1 above, our extended Turing machines
cannot be regarded as physical computers in the narrow sense, 
since they are able to realize
elements of $\Sigma_2$. 

By using ideas of L\'aszl\'o Kalm\'ar, let us try to formulate a more
tangible (and somewhat stronger) version of the above Thesis. Of the many
roles Turing machines play in scientific thinking, let us concentrate on
the following one. Turing machines provide {\it idealized}, abstract
``approximation'' of {\it artificial computing systems}\footnote{Here one 
can think of a ``futuristic'' notion of computer.}. The next version of
the Thesis will say that arbitrary future artificial computing systems
will realize only such functions $f:\N^k\rightarrow\N^m$ which are
Turing-computable (i.e. recursive). To make the meaning of the next
version of the Thesis clear, we ask ourselves what 
artificial computing systems are. The answer is the following.

Any such system presupposes that we fix a physical theory (which is
consistent with our present day knowledge) and on the basis of this theory
we design an artificial system which is capable to associate natural
numbers to natural numbers in some well-defined way. (Here well-defined
means that, in terms of the chosen physical theory, it is clearly
explained  how to give an ``input'' to this system and how to interpret
whether it gave an ``output'' and what this output is.)

But what is an artificial system? Does it have to fit into a box, for
example? If yes, what are the limits of the size of the box? (What happens
if the system uses a futuristic version of, say, Internet? What if this
net grows {\it during} the course of computation in question?) If we do
not want to be ``short-sighted'' we should not suppose that the system
fits into a box (or anything like that).

In view of the above considerations, for the purposes of the present
paper, we propose to identify an artificial computing system $G$ with what
we call here a {\it thought-experiment}\footnote{Assume a physical theory
is fixed. Then by a {\it thought-experiment relative to the fixed physical
theory} we mean a theoretically possible experiment, i.e. an experiment
which can be carefully designed, specified, etc. according to the rules
of the physical theory but for the actual realization of it we might not
have the necessary sources, technical level, enough time, etc.

To illustrate the idea: if the physical theory in question is classical
mechanics then we conjecture that there are no thought-experiments which
would realize a function $f$ which is not Turing-computable.}. 
The definition of when we say that a mathematical {\it function $f$ is
realized by a fixed artificial computing system $G$} (or thought
experiment) follows the same pattern as we defined earlier the concept 
of when a physical computer (in the narrow sense) realizes function
$f$. Therefore we do not repeat that definition.
\vspace{0.1in}

{\bf Definition 4}. We regard the above considerations as the
definition of when a mathematical function $f:\N^k\rightarrow\N^m$ is 
{\it realized by an artificial computing system} $G$. $\Diamond$
\vspace{0.1in}

We would like to clarify a bit the sense in which we use the
expression ``thought-experiment'' in Definition 4 above. If $G$ is a
thought-experiment (i.e. artificial computing system), then there is a
fixed physical theory $Th$ associated to $G$ such that using theory
$Th$ one can specify precisely how the thought experiment $G$ should
be carried out. If using $Th$ together with the specification of $G$
someone can prove that $G$ realizes $f$, then we conclude that indeed
$G$ realizes $f$. We note that this does not mean that using $Th$ and
the specification of $G$ we could compute with pencil and paper what
the answer of $G$ will be to a certain input, say 3. We only know that
$G(3)=f(3)$ holds\footnote{On the other hand, if in a ``universe'' $U$, 
the theory $Th$ was true and someone had the resources for carrying the
thought-experiment through, then at the end he would find out the
value of $f$ for any given prespecified input.}.

\noindent Trivially, the class of artificial computing
systems, defined in this way, includes the class of physical
computers (in the narrow sense) 
used to formulate Thesis 1 (cf. \cite{odi} p. 104). Moreover, 
the question naturally arises whether extended Turing machines introduced
above exist in the class of artificial computing systems or not?  
Notice also that a function $f$ which is realizable by an
artificial computing system is computable of the second kind according to
the terminology developed in Section 1. 

Now we are ready to formulate a sharper version of Thesis 1.
\vspace{0.1in}

{\bf Thesis 2} (Church--Kalm\'ar--Turing). {\it Every function
$f:\N^k\rightarrow\N^m$ realizable by an artificial computing system 
gives rise to a relation $R_f\in\Sigma_1$ i.e. every
function realizable by an artificial computing system
is Turing-computable}. 
\vspace{0.1in}

\noindent Or, trivially re-formulated, we can state:
\vspace{0.1in}

{\bf Thesis 2'} (Church--Kalm\'ar--Turing). {\it Every function
$f:\N^k\rightarrow\N^m$ realizable by a thought-experiment gives rise to a
relation $R_f\in\Sigma_1$ i.e. every function realizable by a
thought-experiment is Turing-computable.}
\vspace{0.1in}

\noindent Clearly, all versions of the Thesis (i.e. 1-2') presuppose some
physical theory as a background. We will argue that the truth of Theses
2-2' can actually depend on the choice of our background physical theory.

A kind of corollary of the Thesis taken together with G\"odel's Second
Incompleteness Theorem is the following.
\vspace{0.1in}

{\bf Thesis 3}. {\it Assume ZFC set theory is
consistent. Then, necessarily, Humankind, or its Successors, can never
prove or become certain that this is so.}
\vspace{0.1in}

\noindent The above form is a kind of common meta-mathematical
interpretation of G\"odel's Second Incompleteness Theorem. We will argue
that the refutability (or provability) of Thesis 3 can also depend on the
choice of our background physical theory.

For completeness, below we will formulate a further version of the Thesis
which goes off in a different angle called sometimes ``limitations of
human knowledge''. This will be Thesis 4-4'. We may formulate Thesis 4 as
follows. If we suppose that the ``input-output aspect'' of each single 
human problem solving activity is nothing but a finite answer to a
finite question formulated in a language fixed in advance, then one may
declare:
\vspace{0.1in}

{\bf Thesis 4} (Church--Descartes--Turing). {\it Every mental activity of
human beings realizes Turing-computable functions}.
\vspace{0.1in}

\noindent This idea can be traced back to {\it Descartes}. If we
accept psychological materialism in the form that every mental product
of a human being is completely determined by his brain the above Thesis
can be re-formulated as
\vspace{0.1in}

{\bf Thesis 4'} (Church--Descartes--Turing). {\it The human brain realizes
Turing-computable functions}.
\vspace{0.1in}

We included Theses 4-4' only for completeness, in our investigations
we will concentrate on Theses 2-3. Our reason for formulating so many
versions of the Thesis is that for {\it each one} of Theses 2-3 we will
argue that  they admit counterexamples if we work in classical  general
relativity theory\footnote{For most of these versions  our main
point is not so much refuting the Thesis but instead is showing that the
Thesis is not independent of the background physical theory. Our main
message is that the theories of computability and meta-mathematics can be
better developed if we take into account the current state of
theoretical physics. To be more blunt: we would like to show that it is
not ``healthy'' to regard and develop these theories as being completely
disjoint and isolated from theoretical physics. In other words what we are
arguing for is the ``unity of science''.}. So, if the reader is
interested in {\it any} one of Theses 2-3 then he can read the rest of
this paper with that version of the Thesis in mind. For definiteness,
we will always formulate our statements to attack Thesis 3.
 
In the following section we try to construct an artificial computing
system based on the ordinary theory of Turing machines and classical general
relativity which is supposed to be able to realize non-Turing-computable
i.e. non-recursive functions. These machines are also counterexamples for
versions 2-3 of  the Church--Kalm\'ar--Turing Theses formulated above.
The basic idea is essentially the same as that of Malament, Hogarth
\cite{hog1} and Pitowsky \cite{pit}; it is summarized by Earman (see
Chapter 4 of \cite{ear}). Moreover we will see that, our
thought-experiment, which is a modified version of the one
constructed in \cite{ear}, is free of the problems listed by Earman and
Pitowsky.

It would be interesting to see which level of the arithmetical hierarchy
can be made ``computable'' by using classical general relativity theory;
and what is the ``price'' of going further up in the hierarchy. That
is, what extra assumptions do we need to make (if any) if we want to make
a higher level of the hierarchy to become ``computable''. The complexity
classes in the analytical hierarchy are denoted by $\Sigma_n^k$ and
$\Pi_n^k$ ($k,n\in\N$). For any of these functions the question whether it
can be made ``computable'' admits a precise, unambiguous formulation {\it
because} all these functions (in $\Sigma_n^k$ etc.) are {\it definable} in
the language of set theory (and even in the higher-order logic language of
arithmetics $\langle\N ,0,1,+,*\rangle$). So, one can write up the
arithmetical definition of the function $f$ and one can ask if there is a
thought-experiment realizing precisely this function.

We note, however, that non-computable functions necessarily remain even if
one uses relativistic (or other) powerful phenomena to compute more and
more complicated functions. The reason for this is a simple
cardinality argument: any thought-experiment can be expressed as a
finite sequence of (English) sentences, therefore there are countably
many thought-experiments only. It follows, that only countably many
functions can be realized by a thought-experiment. On the other hand,
the cardinality of $\N\rightarrow\N$ type functions is the continuum.
Therefore, there must exist a function, which cannot be realized by a
thought-experiment.

\section{Computers in the Kerr space-time}
In this section we will follow the notation and terminology of \cite{ear}
(see also \cite{haw-ell}, \cite{wal}). By a
space-time we mean a pair $(M,g)$ where $M$ is a smooth, oriented and
time-oriented four-manifold while $g$ is a smooth Lorentzian metric on
$M$ which is a solution to the Einstein's equations with respect to a
physically reasonable matter field represented by a smooth stress-energy
tensor $T$ on $M$ (i.e. $T$ satisfies one of the standard
energy conditions). For the notions concerning general relativity we refer
to \cite{haw-ell},\cite{wal}. The length of an at least once
continuously differentiable time-like curve $\gamma:\R\rightarrow M$ is
the integral
\[\Vert\gamma\Vert=\int\limits_\gamma\dd\gamma =\int\limits_\R
\sqrt{-g(\dot\gamma (\tau ) , \dot\gamma (\tau ))}\:\dd\tau .\]
As usual, we interpret a future-directed, time-like, at least once
continuously differentiable curve $\gamma :\R\rightarrow M$ as a
``world-line'' of an observer moving in $(M,g)$, i.e. ${\rm
im}\gamma\subset M$ is the collection of those events in $M$ which the
observer meets throughout its existence. Moreover $\Vert\gamma\Vert$, the
length of the world-line, is thought of as the proper time measured by the
observer $\gamma$ from its beginning of existence to its end. This can be
finite or infinite depending on the curve and the geometrical structure of
the space-time characterized by the metric $g$. Now we introduce an
important class of space-times related with our subject. Consider a point
(event) $q\in M$. The set of all points
\[J^-(q):=\{ x\in M\:\vert\:\mbox{there is a future-directed
non-space-like continuous curve joining $x$ with $q$}\}\]
is called the {\it causal past} of the event $q$ (the causal future is
defined similarly). Intuitively, $J^-(q)$ consists of those events $x\in M$
from which one can ``travel'' to $q$ without exceeding locally the speed
of light i.e. by an ``allowed'' motion.
\vspace{0.1in}
 
{\bf Definition 5.} A space-time $(M,g)$ is called a {\it
Malament--Hogarth space-time} if there is a future-directed time-like
half-curve $\gamma_P :\R^+\rightarrow M$ such that $\Vert\gamma_P\Vert
=\infty$ and there is a point $p\in M$ satisfying ${\rm im}\gamma_P\subset
J^-(p)$. The event $p\in M$ is called a {\it Malament--Hogarth event}.
$\Diamond$
\vspace{0.1in}

\noindent Note that if $(M,g)$ is a Malament--Hogarth space-time, then
there is a future-directed time-like curve $\gamma_O :[a,b]\rightarrow M$
from a point $q\in J^-(p)$ to $p$ satisfying $\Vert\gamma_O\Vert<\infty$.
The point $q\in M$ can be chosen to lie in the causal future of the past
endpoint of $\gamma_P$. Below we will discuss if such space-times are
physically reasonable or not.

Consider a Turing machine realized by a physical computer $P$
moving along the curve $\gamma_P$ of {\it infinite} proper time. Hence the
physical computer (identified with $\gamma_P$) can perform arbitrarily long
calculations. Being $(M,g)$ a Malament--Hogarth space-time, there is
an observer following the curve $\gamma_O$ (hence denoted by
$\gamma_O)$ of {\it finite} proper time such that he touches the
Malament--Hogarth event $p\in M$ in finite proper time. But by definition 
${\rm im}\gamma_P\subset J^-(p)$ hence in $p$ he can receive the answer
for a {\it yes or no question} as the result of an {\it arbitrarily long} 
calculation carried out by the physical computer $\gamma_P$ since
it can send a light beam to $\gamma_O$ at arbitrarily late proper
time. Clearly the pair $(\gamma_P, \gamma_O)$ is an {\it artificial
computing system} $G$ with respect to classical general relativity theory
since it is a correct thought-experiment within the framework of this
theory. Hence $G:=(\gamma_P, \gamma_O)$ carries out a {\it computation of 
the second kind}. In this moment, it is not clear what kind of
space-time $M$ is and what time-like curves $\gamma_P$ and $\gamma_O$ are. For
instance, it is possible that the acceleration along one curve is
unbounded making the idea physically unreasonable \cite{pit}. A very
concrete, physically reasonable realization of this device in the case of
the Kerr space-time will be explained soon.
 
Imagine the following situation, as an example. $\gamma_P$ is asked to
check all theorems of our usual set theory (ZFC) in order to check
consistency of mathematics. This task can be carried out by $\gamma_P$
since its world line has infinite proper time. If $\gamma_P$ finds a
contradiction, it can send a message (for example a light beam) to
$\gamma_O$. Hence if $\gamma_O$
receives a signal from $\gamma_P$ {\it before} the Malament--Hogarth event
$p$ he can be sure that ZFC set theory is not consistent. On the other
hand, if $\gamma_O$  does not receive a signal before  $p$ then, {\it 
after} $p$, $\gamma_O$ can conclude that ZFC set theory is consistent.
Note that $\gamma_O$ having finite proper time between the events
$\gamma_O(a)=q$ (starting with the experiment) and $\gamma_O(b)=p$
(touching the Malament--Hogarth event), he can be sure about the
consistency of ZFC set theory in finite (possibly very short) time. This
contradicts Thesis 3 above.

At this point we may ask if Malament--Hogarth space-times are reasonable
physically or not. Most examples are very artificial but it is quite
surprising that among these space-times one can recognize the {\it anti-de
Sitter space-time}, which is a solution to the vacuum Einstein's 
equations with negative cosmological constant and is in the focus of
recent investigations in theoretical physics; the {\it
Reissner--Nordstr\"om space-time} describing
a spherically symmetric black hole of small electric charge and the {\it
Kerr--Newman space-time} representing a slowly rotating black hole of
small electric charge.
For a description of these space-times see \cite{haw-ell} as a standard
reference. In what follows we are going to focus our attention to the
Kerr space-time because in light of the celebrated black hole
uniqueness theorem (see \cite{haw-ell}, or for an
overview \cite{wal} while a short new proof was presented by Mazur
\cite{maz}) 
this space-time is the only candidate for the late-time evolution of a
collapsed rotating star. Hence existence of Kerr black holes in the
Universe is physically very reasonable even in our neighbourhood.
For instance, a candidate for such a black hole is the
supermassive compact object in the center of the Milky Way; this
question can be decided in the next few decades \cite{mel}. In this
context it is remarkable that this space-time possesses the
Malament--Hogarth property.

Now we would like to construct the artificial computing system
$G=(\gamma_P,\gamma_O)$ as a correct thought-experiment in the case of the
vacuum Kerr space-time $(M,g)$. This means that we have to describe the
time-like curves $\gamma_O$ and $\gamma_P$ around a slowly rotating black
hole of zero electric charge. To do this, we will follow \cite{o'n}. Using
Boyer--Lindquist coordinates $(t,r,\vartheta, \varphi )$, the Kerr metric
$g$ with parameters $m>0$ (mass) and $a$ (angular momentum per unit mass)
locally takes the shape
(see \cite{haw-ell}\cite{o'n}\cite{wal})
\[\dd s^2=-\left( 1-{2mr\over\Sigma}\right)\dd
t^2-{2mra\sin^2\vartheta\over\Sigma}\dd
t\dd\varphi+{\Sigma\over\Delta}\dd r^2+\Sigma\dd\vartheta^2+\left(
r^2+a^2+{2mra^2\sin^2\vartheta\over\Sigma}\right)
\sin^2\vartheta\dd\varphi^2\]
where $\Sigma (r,\vartheta ) = r^2+a^2\cos^2\vartheta$ and $\Delta (r)
=r^2-2mr+a^2$. We
choose the underlying manifold $M$ to be a smooth four-manifold which
can carry the maximal analytical extension of the metric\footnote{This
determines the range of the values of $t, r,\vartheta ,\varphi $, see
\cite{haw-ell}\cite{o'n}.}. This metric possesses two
Killing fields, namely $\partial/\partial t$ and
$\partial/\partial\varphi$ corresponding
to time-translations and rotations around the ``axis'' of the black hole,
respectively. The singularity is given by the equation $\Sigma (r,
\vartheta ) =0$ and has ring-shape while the event horizons
are characterized by the real roots $r_\pm$ to the equation $\Delta
(r)=0$:
\[r_\pm =m\pm\sqrt{m^2-a^2}.\]
Note that this equation has real roots only if $\vert a\vert\leq m$,
i.e. in the case of ``slowly rotating'' black holes. We restrict
ourself to the non-extremal case $\vert a\vert <m$. 

Assume a future-directed time-like geodesic $\gamma : \R^+\rightarrow M$
is given, describing the free motion of a point-like particle of unit
mass. In the above coordinate system this curve locally is given by the
four functions $\gamma (\tau )=(t(\tau ), r(\tau ), \vartheta (\tau ),
\varphi (\tau ))$ satisfying the well-known second order geodesic
equations. We can identify such a curve uniquely by fixing the initial
position and velocity $(\gamma (0),\dot\gamma (0))$ where dot means
differentiation with respect to the affine parameter
$\tau\in\R^+$. However, if $\gamma (0)$ is not on the axis
of the black hole, then by Lemma 4.2.5 of \cite{o'n} we can use the data
$(\gamma (0),{\rm sgn}\:\dot r(0),{\rm sgn}\:\dot\vartheta (0), q, E, L,
Q)$ to fix the geodesic $\gamma$ as well\footnote{Here sgn is the sign of
a real number.}. The quantities $(q, E,L, Q)$ are the ``first integrals''
of the geodesic motion, i.e. these quantities are constant along the
geodesic curve. Here $q:=g(\dot\gamma ,\dot\gamma )$ is equal to $-1$
since $\gamma$ is time-like and the point particle is of unit
mass. $E:=-g(\dot\gamma , \partial/\partial t)$ is the total energy of
the particle measured by a distant observer, $L:=g(\dot\gamma
,\partial/\partial\varphi )$ is the angular momentum of the particle with
respect to the ``axis'' of the black hole given by points
satisfying $\vartheta =0, \pi$.
The constant $Q$ is called the {\it Carter-constant} and is characterized
by the system of ordinary differential equations (see Section 4.2 of
\cite{o'n})
\[\Sigma^4(r, \vartheta )\dot r^2 =-\Delta (r) (r^2 +Q
+(L-aE)^2)+(r^2+a^2)E-aL,\]
\[\Sigma^4 (r\, \vartheta )\dot\vartheta^2=Q+(L-aE)^2-a^2\cos^2\vartheta
-{L\over\sin^2\vartheta}-aE.\]
A remarkable observation of Carter shows that $Q$ is constant along a
Kerr-geodesic (see Theorem 4.2.2 of \cite{o'n}) and characterizes
Kerr-geodesics in a simple way whether they hit or not the ring
singularity. 

First, we consider the freely falling observer $\gamma_O:
[0,\tau_-]\rightarrow M$. Choose a particular point $q\in M$ somewhere
``outside'' the black hole, not lying on the axis and let $\gamma_O(0):=q$. Let
${\rm sgn}\:\dot r_O(0)=-1$, while ${\rm sgn}\:\dot\vartheta_O(0)=\pm 1$
arbitrary and take $0<E_O$, $\vert L_O\vert <2mE_Or_+/a$. These data
provide for a (particle-like) observer moving along $\gamma_O$ to enter
the Kerr black hole, i.e. to cross the outer event horizon.
Moreover, if we take $Q_O\not =0$ then by Corollary 4.5.1 of
\cite{o'n} $\gamma_O$ does not hit the singularity $\Sigma =0$ of
the black hole. Furthermore, if we fix $E_O^2\geq 1$ then the passenger
has enough energy to escape some infinite, asymptotically flat region
of $M$ again (see Proposition 4.8.1 of \cite{o'n}) particularly he crosses
the inner horizon as well. Finally, if we choose the angular momentum
$L_O$ of the geodesic $\gamma_O$ carefully, namely
\[{2mE_Or_-\over a}<L_O<{2mE_Or_+\over a}\] 
(in particular this gives $0<L_O$, showing $\gamma_O$ cannot be an 
axial geodesic since in that case $L=0$), then $\gamma_O$ hits
the inner horizon in a Malament--Hogarth event (see Fig. 4.19 of \cite{o'n}). 
It is worth mentioning at this point that such an orbit does not exist
for non-rotating ($a=0$) i.e. Schwarzschild black holes. The above type of
geodesics are called ``time-like long flyby orbits of type B'' and are
examined on pp. 245--247 of \cite{o'n}. The Malament--Hogarth event is
characterized by the equation $r_O(\tau_-)=r_-$. Clearly, $\tau_-$ is
finite since $\gamma_O$ reaches the inner horizon under the above
conditions, hence
\[\Vert\gamma_O\Vert =\int\limits_0^{\tau_-}\sqrt{-g(\dot\gamma_O(\tau 
), \dot\gamma_O(\tau ))}\:\dd\tau =\int\limits_0^{\tau_-}\dd\tau
=\tau_-<\infty .\]

The case of the physical computer is very simple. We may assume the
initial data are $\gamma_P(0)=\gamma_O(0)=q$ (the observer $\gamma_O$ and
the computer $\gamma_P$ start from the same point) and take $\gamma_P:
\R^+\rightarrow M$ to be a geodesic corresponding to a stable circular
orbit in the equatorial plane of the Kerr black hole. This implies ${\rm
sgn}\:\dot r_P(0)=0$, ${\rm sgn}\:\dot\vartheta_P(0)=0$, $Q_P=0$ 
and $E_P>0$, $E_P^2<1$. We can calculate the
radius of the circular orbit of $\gamma_P$ by Lemma 4.14.9 while the
corresponding angular momentum $L_P$ can be determined via Corollary 
4.14.8 of \cite{o'n} (the concrete values are not interesting for us in
this moment). Trivially, $\Vert\gamma_P\Vert=\infty$. 

This arrangement shows that, since both $\gamma_P$ and $\gamma_O$ move
along geodesics, their acceleration is constantly zero, i.e. remains
bounded throughout their existence. A three-dimensional picture of the
machine is shown on Figure 1.
\vspace{0.1in}

\centerline{\psfig{figure=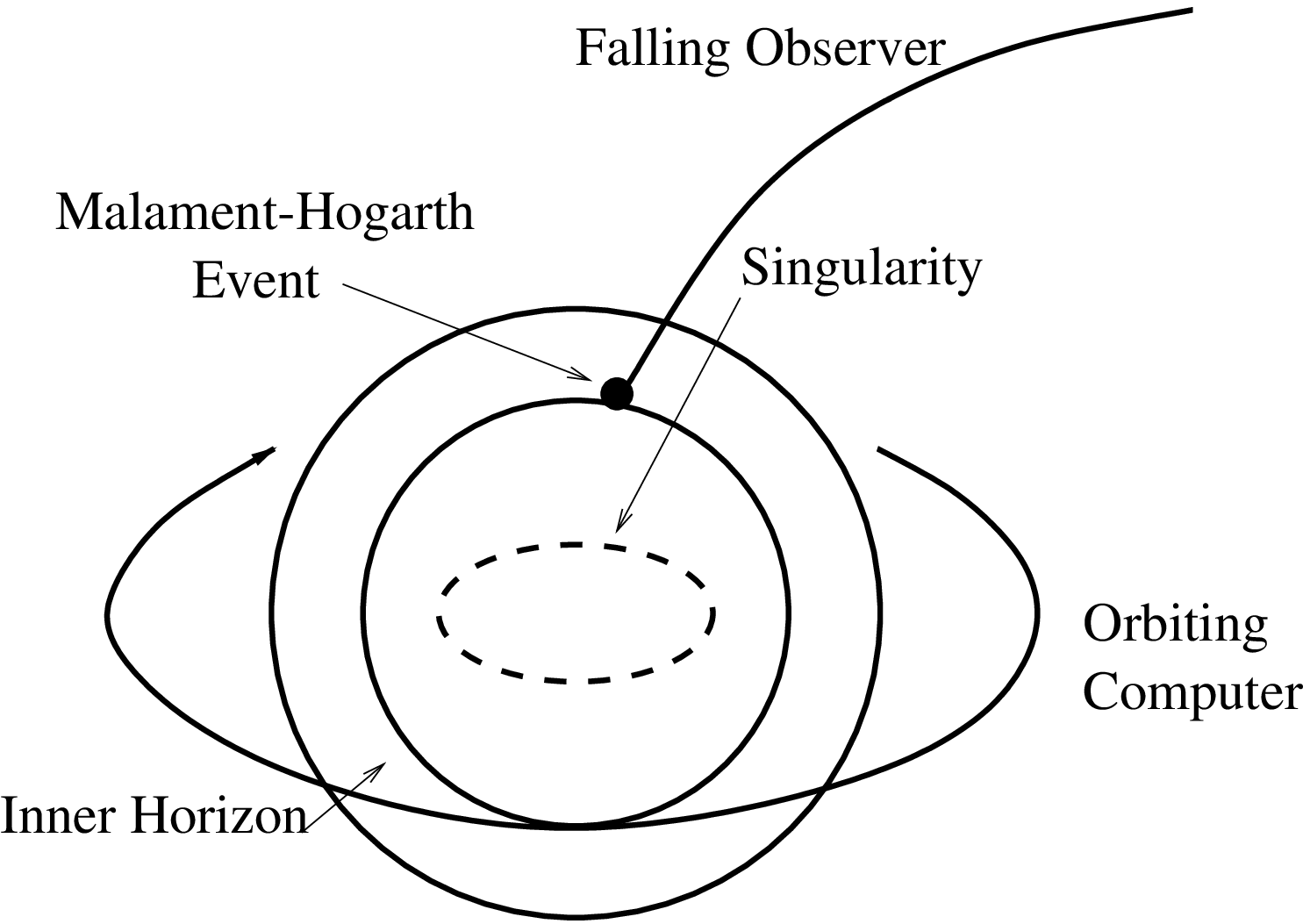,height=2in}}
\vspace{0.1in}

\centerline{{\bf Figure 1.} The three-dimensional picture of the
device $G=(\gamma_P, \gamma_O)$.}
\vspace{0.1in}

\noindent It is worth presenting a four-dimensional space-time diagram of
$G=(\gamma_P, \gamma_O)$ as well. Such diagrams
are called Penrose diagrams and show the whole development of the system.
\vspace{0.1in}

\centerline{\psfig{figure=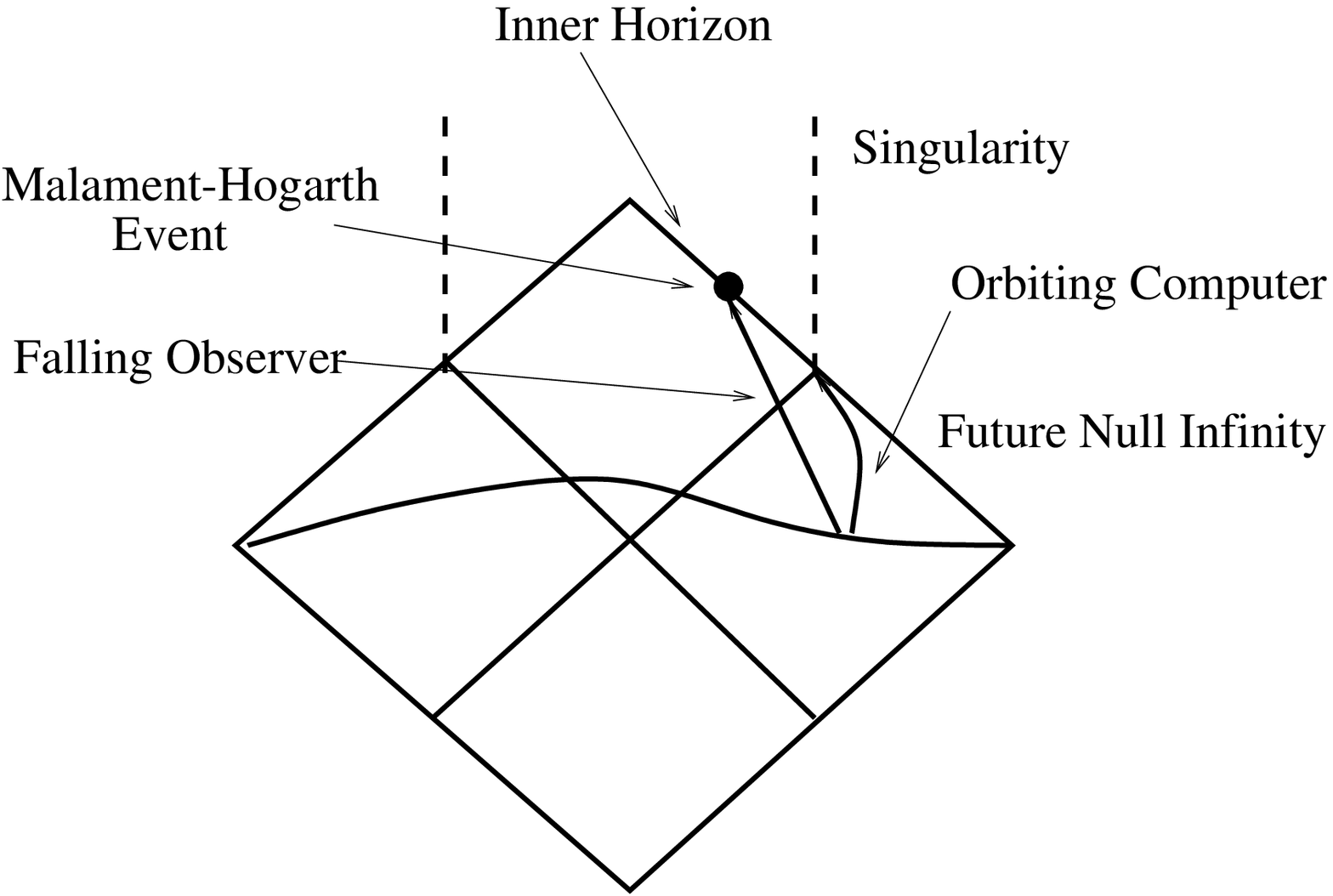,height=2in}}
\vspace{0.1in}

\centerline{{\bf Figure 2.} The Penrose diagram picture of the
device $G=(\gamma_P, \gamma_O)$.}
\vspace{0.1in}

\noindent We can see that in the case of Kerr space-time the
Malament--Hogarth event appears for $\gamma_O$ when he touches the inner
horizon of the Kerr black hole (in a finite proper time, of course). As it
is well known \cite{haw-ell}\cite{wal} the inner horizon of the Kerr black
hole is a Cauchy horizon for outer observers showing that this space-time
fails to be globally hyperbolic. Later we will see that this is a
general property of Malament--Hogarth space-times. Although after crossing
the inner horizon the predictability of the fate of $\gamma_O$ breaks
down, it seems he can avoid the encounter with the final destroying
singularity in the stomach of the Kerr black hole as a consequence of the
ring-like shape of the singularity.      

Now that the Kerr-orbits of the falling traveler $\gamma_O$ and the
orbiting computer $\gamma_P$ are determined, let us turn our attention
to the communication between them by fixing a simple coding
system. For sake of definiteness, assume we want to attack Thesis 3.
Consequently we have to derive all the theorems $\phi_1,\phi_2,\dots$ of
ZFC set theory and check if there exists a theorem, say $\phi_i$, which
coincides with the formula FALSE or equivalently with $x\not= x$. Then
$\gamma_O$ and $\gamma_P$ choose a Turing machine $T$ which enumerates all
the theorems of ZFC. In this way $T$ realizes a function
$f_T:\N\rightarrow\{\mbox{Formulas of ZFC}\}$ such that im$f_T$ is exactly
the set of theorems in ZFC (it is easy to find such a $T$). Now,
$\gamma_O$ and $\gamma_P$ agree on using the same choice of $T$. Then
$\gamma_O$ departs for the Kerr black hole (taking a copy of $T$ with him) 
while $\gamma_P$ keeps on executing the following simple algorithm.
\[\begin{array}{ll}
\mbox{A.} & i:=0\\
\mbox{B.} &\mbox{Derive theorem $f_T(i)$ from ZFC set theory}\\
\mbox{C.}& \mbox{Check if $f_T(i)=$ FALSE}\\
\mbox{D.}& \mbox{If yes, send a signal to $\gamma_O$}\\
\mbox{E.}& \mbox{If no, let $i:=i+1$ and go to B}\\
\end{array}\]
Suppose that ZFC is inconsistent. Then $\gamma_P$ will find the first
$i\in\N$ for which $f_T(i)=$FALSE. Suppose the proper time needed for
$\gamma_P$ to find this $i$ was $\tau_P^i$ (the
experiment started at $\tau_O=\tau_P=0$). Let us mention that for anyone
who has a copy of $T$ and knows the speed of $\gamma_P$'s implementation
of $T$, the number $i$ is computable from $\tau_P^i$. 

Since $\gamma_P$ knows when it is sending the signal and it knows
$\gamma_O$'s plans, $\gamma_P$ can compute how much time $\gamma_O$ will
have for receiving the coded signal and can also compute the expectable
blueshift of the signal (see Section 4). So $\gamma_P$ can make
compensations for these effects (to the extent theoretically
possible). 

Now, $\gamma_P$ sends off a signal. $\gamma_O$ receives it before the
Malament--Hogarth event $p$ and measures the time
$\tau_O^i$ (according to his own clock) when the signal arrived (we will
return soon to the question of measurement of this signal). By knowing the
time $\tau_O^i$ and by using general relativity theory, $\gamma_O$ can
compute the time $\tau_P^i$ hence the number $i$. Then $\gamma_O$ computes 
$f_T(i)$ and checks if it is the formula FALSE. If yes, he knows that ZFC
is inconsistent. If not, then $\gamma_O$ received a fake 
signal\footnote{As $\gamma_O$ approaches the
Malament--Hogarth event which lies on the inner horizon of the Kerr black
hole i.e. on a Cauchy horizon of the Kerr space-time, it is more and more
difficult to decide whether a light beam came from $\gamma_P$ or a
possible past singularity, see \cite{ear}, p. 118. Consequently
receiving fake signals cannot be {\it a priori} excluded.}.

To keep the number of possible fake signals at minimum, we may assume 
that $\gamma_P$ will not send a simple light beam only but uses some
modulation or coding (some Morse-type sequence of ``long'' and ``short''
impulses, for instance) to make its signal much more unique. We emphasize
that this modulation, or coding is also fixed once and for all in advance
between $\gamma_O$ and $\gamma_P$.

If $\gamma_O$ does receive a signal before the Malament--Hogarth
event, then he checks if the relevant theorems in ZFC are consistent
or not. If yes, then $\gamma_O$ concludes that what he received was a fake
signal.

If he did not receive any other signal by $p$, then he concludes that ZFC
is consistent. If he received the prearranged coded signal at some
different time, say $\tau_O'$, too, then he goes through the above
checking procedure for deciding whether this second signal is fake or
not. We assume that $\gamma_O$ and $\gamma_P$ agree on a sufficiently
complicated and long code to minimize the chance for fake
signals. Further, by the nature of the possible origin (or cause) of a
fake signal and by taking into account that on $\gamma_O$'s clock only
finite time goes by between $\gamma_O$'s departure and his arrival at the
Malament--Hogarth event $p$, we can expect that $\gamma_O$ will receive
only finitely many fake signals (before reaching $p$, of course). 
Consequently $\gamma_O$ has to check only a finite number
of signals and after that he will know whether or not ZFC is consistent.

Let us briefly return to the possible imprecision of $\gamma_O$'s
measuring $\tau^i_O$. Suppose $\gamma_O$ knows only that the signal
arrived between $\tau^i_O$ and $\tau^i_O+\varepsilon_O$ (with
$\tau^i_O+\varepsilon_O$ being before the Malament--Hogarth event). Then
he will calculate that it was sent between $\tau^i_P$ and
$\tau^i_P+\varepsilon_P$. But only finitely many theorems were checked by
$\gamma_P$ within this interval, consequently $\gamma_O$ can corrigate
this uncertainty with finite calculations only (i.e. by checking the
falsity of finitely many theorems from ZFC only).
 
Hence, by assuming the ability of time measurement of arbitrary accuracy
(which is always possible in classical physics, but see remarks in Section
4), the arrangement $G$ provides a thought-experiment, consistent with
classical general relativity, contradicting Thesis 3. Having designed an
artificial computing system which checks consistency of ZFC, we now turn
to seeing what other jobs can be done with similar artificial computing
systems. Let us return in general to Theses 2-3 formulated in Section
2. As we said in that section, first we have to assume a physical
theory. Let this theory be the classical general relativity. Next, let us
suppose that the observer $\gamma_O$ wants to decide a $\Sigma_1$-set of
$\N$ which is not $\Pi_1$ i.e. recursively enumerable but
non-decidable. The above considerations can be used by $\gamma_O$ for
designing a thought-experiment that is, an artificial computing system
$G=(\gamma_P, \gamma_O)$ which will help him to decide such a set.
\vspace{0.1in}

{\bf Definition 6.} Let $R\subseteq\N^m$ be a relation. We say that an
artificial computing system (or thought-experiment) $G$ {\it decides} $R$
if and only if $G$ realizes the characteristic function
$\chi_R:\N^m\rightarrow\{ 0,1\}$. $\Diamond$
\vspace{0.1in}

From now on, we will call $G=(\gamma_O,\gamma_P)$ a {\it relativistic
computer}, indicating that this is a special artificial computing system
i.e. thought-experiment. Now we are in a position to state:
\vspace{0.1in}

{\bf Proposition 1.} (i) {\it Let $R\in\Sigma_1$ be a relation with
possibly $R\not\in\Pi_1$ i.e. recursively enumerable but possibly
non-decidable. Then there is a relativistic computer $G=(\gamma_P,
\gamma_O)$ which decides $R$.}

(ii) {\it There exist infinitely many
relations $R\in\Sigma_1\setminus\Pi_1$. Hence there are infinitely many
Turing-undecidable relations which are decidable by some $G=(\gamma_P
,\gamma_O)$ as in} (i) {\it above.}
\vspace{0.1in}

{\it Proof.} (ii) This is well known (c.f. \cite{odi}). An example 
is if we take $R$ to be the set of valid theorems of first-order logic. 

(i) Let $R\in\Sigma_1$. Then $R$ is recursively enumerable i.e. there is a
Turing machine $T$ which enumerates $R$. (In other words, $T$ realizes a
surjective function $f_T: \N\rightarrow R$ with im$f_T=R$.)

Now, we design the relativistic computer $G$ which, we claim, can
decide $R$. To test this claim, the ``opponent'' chooses a random element
$(x_1,\dots ,x_k)\in\N^k$ and gives it to $G$ for deciding whether or not
$(x_1,\dots x_k)\in R$. In the initial state of their computation,
$\gamma_O$ and $\gamma_P$ are sitting together, making plans about how to
decide this question. $\gamma_P$ receives the task of using $T$ to
enumerate the elements of $R$ and checking whether $(x_1,\dots
,x_k)\in\N^k$ shows up during this enumeration. That is, 
$\gamma_P$ executes the program 
\[\begin{array}{ll}
\mbox{A.} & i:=0\\
\mbox{B.} &\mbox{If $f_T(i)=(x_1,\dots ,x_k)$ then send a signal to
$\gamma_O$ and go to D}\\
\mbox{C.}& \mbox{$i:=i+1$ and go to B}\\
\mbox{D.}& \mbox{Make sure that the signal for $\gamma_O$ is adequately
coded, prepared etc.}\\
&\mbox{Make other planned actions to ensure that
$\gamma_O$ receives the signal. STOP}\\ 
\end{array}\]
The rest of the preparations $\gamma_P$ and $\gamma_O$ make are
exactly the same as was the case of the relativistic computer
$G$ described above Proposition 1 for refuting Thesis 3. (So here again
they rely on precise measurment of time to rule out fake signals, and
again $\gamma_O$ takes the Turing machine $T$ with him such that he can
compute $f_T(i)$ for any fixed $i$).

After the Malament--Hogarth event $p$, $\gamma_O$ will be able to decide
whether the input $(x_1,\dots ,x_k)$ received from the ``opponent'' is in
$R$ because if he received a signal (before $p$) and he
(successfully) checked the signal for correctness in the above outlined
way, then he knows $(x_1,\dots ,x_k)\in R$. Otherwise he knows
$(x_1,\dots ,x_k)\notin R$. We finished the proof. $\Diamond$  
\vspace{0.1in}

{\bf Corollary 1.} {\it There are infinitely many functions
$f:\N\rightarrow\N$ such that}

(i) {\it f is realized by a relativistic computer
$G=(\gamma_P,\gamma_O)$};

(ii) {\it $f$ is non-Turing computable.}
\vspace{0.1in}

{\it Proof.} Let $R\in \Sigma_1\setminus\Pi_1$. It is known that
there are infinitely many such sets, cf. e.g. \cite{odi}. Let $f:=\chi_R$
be the characteristic function of $R$. Then $f:\N\rightarrow\{
0,1\}$ is non-Turing computable because $R$ is undecidable by
$R\notin\Pi_1$. Let $G:=(\gamma_P,\gamma_O)$ be the relativistic
computer deciding $R$. This exists by Proposition 1. Let $G'$ be the same
but instead of ``yes'' or ``no'' let it give as an output 1 or 0. Then
$G'$ realizes $f$. $\Diamond$
\vspace{0.1in}

\noindent Below we will prove stronger theorems. By Proposition 1,
relativistic computers can decide any undecidable but recursively
enumerable relations $R\in \Sigma_1\setminus\Pi_1$. It is natural
to ask whether harder sets of natural numbers become decidable if we
switch to relativistic computers. The next proposition says that the
answer is in the affirmative. 
\vspace{0.1in} 

{\bf Proposition 2}. {\it Let $n>0$. There are infinitely many relations
$R\in\Sigma_2\setminus (\Sigma_1\cup\Pi_1)$, $R\subseteq\N^n$ such that some
relativistic computer decides $R$.} 
\vspace{0.1in}

{\it Proof.} Let $H\in \Sigma_1\setminus \Pi_1$ be
arbitrary. Define 
\[R:=\{ (x,1)\:\vert\: x\in H\}\cup\{ (y,0)\:\vert\: y\in\overline{H}\}
,\]
where $\overline{H}=\N^n\setminus H$. That is, 
\[R=(H\times\{ 1\})\cup (\overline{H}\times\{ 0\} ).\]

(i) $R\notin\Sigma_1$ because we cannot enumerate its second part  
$\overline{H}\times\{ 0\}$ and $R\notin\Pi_1$ because we cannot enumerate
the complement of its first part $H\times\{ 
1\}$. (Hint: $R\in\Sigma_1\Rightarrow$ we can enumerate $R$ $\Rightarrow$
we can enumerate those elements of $R$ which end with 0 $\Rightarrow$ we
can enumerate $\overline{H}\times\{ 0\}$ $\Rightarrow$ we can
enumerate $\overline{H}$.) It can be seen that $R\in\Sigma_2$ (by
$\overline{H}\in\Pi_1$). 

(ii) By Proposition 1 there is a relativistic computer $G$ deciding $H$.

The new $G'$ deciding $R$ does the following: 
If it receives an input $(x,k)$ and
if $k>1$, then $G'$ answers ``no''. Assume $k\leq 1$. Then $G'$ asks
$G=(\gamma_P, \gamma_O)$ to decide whether $x\in H$. If $k=1$ then $G'$
prints out the same answer as $G$. If $k=0$ then $G'$ prints out the
negation of the answer of $G$.

Clearly, $G'$ is a relativistic computer deciding  $R\in\Sigma_2\setminus
(\Sigma_1\cup\Pi_1)$. $\Diamond$ 
\vspace{0.1in}

\noindent Since $R\notin\Sigma_1\cup\Pi_1$, our new computer $G'$
constructed in the proof of Proposition 2 decides sets harder than
{\it recursively enumerable sets} and {\it complements of recursively
enumerable ones}. This means that we can ``climb higher'' with one extra
degree of unsolvability with Proposition 2. We have the following
corollary, immediate from Proposition 2. 
\vspace{0.1in}

{\bf Corollary 2.} {\it There are infinitely many $\Sigma_2\setminus
(\Sigma_1\cup\Pi_1)$ functions $f:\N\rightarrow\N$ realizable by
relativistic computers. (Of course these functions are non-Turing
computable).} $\Diamond$
\vspace{0.1in} 

\noindent We note that the simplest examples of
$R\in\Sigma_2\setminus\Sigma_1$ relations are the characteristic functions 
$\chi_H$ of relations $H\in\Sigma_1\setminus\Pi_1$. We claim that
relations decided relativistically by Proposition 2 cannot be obtained in
this way. Therefore by Proposition 2 we can decide $\Sigma_2$-relations
which are strictly more complex (i.e. harder) than the simplest
examples for $R\in\Sigma_2\setminus\Sigma_1$.

Let us ask if we can decide even harder sets than in Proposition 2. Each
relation decided by Proposition 2 can be regarded as a disjoint union of a
$\Sigma_1$-set and a $\Pi_1$-set. In the next proposition we will decide   
relations in $\Sigma_2\setminus (\Sigma_1\cup\Pi_1)$ which cannot be
obtained as such disjoint unions. In some sense this means that we can
decide even broader spectrum of hard relations. 
\vspace{0.1in}

{\bf Proposition 3}. {\it Let $n>0$. There are infinitely many relations 
$R\in\Sigma_2\setminus (\Sigma_1\cup\Pi_1)$, $R\subseteq\N^n$ such that}

(i) {\it $R$ cannot be obtained as a disjoint union of finitely many
$\Sigma_1$ and $\Pi_1$ relations;}

(ii) {\it $R$ is decidable by a relativistic computer
$G=(\gamma_O,\gamma_P)$.}
\vspace{0.1in}

{\it Proof.} Let $H\in \Sigma_1\setminus\Pi_1$ be arbitrary, $n>0$. Define 
\[X_H:=\{ (a,b)\:\vert\: a\in H \mbox{ and }b\in\overline{H}\}.\]
Let $\chi_{X_H}=:f:\N^{2n}\rightarrow \{ 0,1\}$ be the characteristic
function of $X_H$. Then $R_f\subseteq\N^{2n}\times\{
0,1\}\subset\N^{2n+1}$ is a $2n+1$-ary relation.

(i) To decide $R_f$, our $G=(\gamma_O,\gamma_P)$ is similar to the one that
was before but now $\gamma_P$ can send {\it two} different kinds of
signals to $\gamma_O$, say $S_a$ and $S_b$. The input for $G$ is of the
form $(a,b,k)=((a,b),k)$ (where $k$ refers to the $(2n+1)$-th
component of $R_f$). The case distinction between
$k>1$ and $k\leq 1$ is similar to that in the proof of Proposition 2.
If $k>1$ then $\gamma_O$ automatically prints ``no''.  

Assume $k=1$. $\gamma_P$ does the following: It starts searching for $a$
in $H$. If it finds $a\in H$ then sends out $S_a$ to $\gamma_O$ and starts     
a search for $b\in H$. If it finds $b$, then sends $S_b$. Now
$\gamma_O$ does the following: If he receives no signal, then prints out
``no''. If he receives $S_a$ and no $S_b$ then prints ``yes''. If he
receives  $S_b$ and no $S_a$ then prints ``no''. Finally, if he receives
both an $S_a$ and $S_b$, then he prints ``no''.

Assume $k=0$. Then $\gamma_P$ starts two parallel processes $P_a$ and
$P_b$. If $P_a$ finds $a\in H$ it sends off a $S_a$ while if $P_b$ finds
$b\in H$ it sends $S_b$. If $\gamma_O$ receives no signal, 
he prints ``yes''. If he receives $S_a$ but no $S_b$ then prints
``no''. If he receives an $S_b$ then prints ``yes''
(independently of other possibly received signals).

(ii) In connection with $R_f$ not being a disjoint union of a $\Sigma_1$ and
a $\Pi_1$ set, we note only the following. Let 
\[R_i:=\{ (a,b,i)\:\vert\: (a,b,i)\in R_f\} .\]
Then $R_1=R\times\{ 1\}$ with $R=T_H$. 
So, in some sense, the ``complexity'' of $R_1$
is determined by the ``complexity'' of $R$. But $R$ was of the form 
$R=\{ (a,b)\:\vert\: a\in H\mbox{ and }b\notin H\}$ with
$H\in\Sigma_1\setminus\Pi_1$. So clearly $R\notin\Sigma_1$ because
of the ``$b$ part'' and  $R\notin\Pi_1$ because of the ``$a$ part''. To
save space, we omit the rest of the proof, since the present proposition
is not of a central importance. $\Diamond$
\vspace{0.1in}

\noindent By Proposition 3 above, relativistic computers can solve
problems much harder than the non-Turing computable problem of deciding an
undecidable but recursively enumerable (i.e. $\Sigma_1\setminus\Pi_1$) 
relation.

The next proposition shows that the {\it extended Turing machine}, we
discussed between Definition 3 and Definition 4 in Section 2---which for
any Turing machine $T$ and possible input $(x_1,\dots ,x_k)$ decides
whether $T$ terminates---is also realizable by a relativistic computer
$G$. 
\vspace{0.1in}

{\bf Proposition 4.} {\it There is a relativistic computer $G=(\gamma_P,
\gamma_O)$ which takes as input a program} pr$(T)$ {\it for a Turing
machine $T$ and a possible input $(x_1,\dots ,x_k)$ for $T$. Then $G$
yields output ``diverges'' if $T$ diverges for $(x_1,\dots,x_k)$ or else
``converges with output $(y_1,\dots ,y_l)$'' if $T$ indeed converges for
input $(x_1,\dots ,x_k)$ with output $(y_1,\dots ,y_l)$.}
\vspace{0.1in}

{\it Proof.} $G$ is of the form $(\gamma_P, \gamma_O)$  as
usual. $\gamma_P$ and $\gamma_O$, sitting together, receive as input a
program pr$(T)$ for some $T$ and a possible input $(x_1,\dots ,x_k)$ for
$T$. For simplicity we will write ``$T$'' for pr$(T)$.

Then $\gamma_O$ takes a copy of $T$ and $(x_1,\dots ,x_k)$ with himself
and
starts his journey ``toward the Malament--Hogarth event'' $p\in M$. Then
$\gamma_P$ starts executing $T$ with input $(x_1,\dots ,x_k)$. If $T$
terminates, $\gamma_P$ sends a signal to $\gamma_O$. At (or after) the
Malament--Hogarth event, $\gamma_O$ does the following. If no signal
arrived then prints ``diverges''. If he received a signal from $\gamma_P$
then $\gamma_O$ knows that $T$ converges with
$(x_1,\dots ,x_k)$. Consequently $\gamma_O$ can safely start executing $T$
with $(x_1,\dots ,x_k)$ and he knows that $T$ will terminate in finite
time. When $T$ terminates, then $\gamma_O$ prints out whatever output $T$
yielded. $\Diamond$ 
\vspace{0.1in}

\noindent In light of Propositions 1-4 and Corollaries 1-2 above  we can
decide general $\Sigma_1$-sets and are able to realize hard 
$\Sigma_2$-sets by our relativistic computer $G$ contradicting Theses
2-2'. In our opinion, the above considerations point in the direction that
if we choose classical general relativity as the background physical
theory then Theses 2-3 turn out to be false because they deal with
computability of the second kind. The reason why Thesis 1 cannot be
attacked is the difference between an artificial computing system
(i.e. a thought-experiment in a consistent physical theory) and
a physical computer in the narrow sense: it is possible that our
artificial computing system cannot be realized as a physical computer,
although we remark that the almost-sure existence of large rotating black
holes in galactic nuclei and properties of these black holes (see
below) point towards the effective realizability of our
thought-experiment, i.e. towards the possible violation of Thesis 1, too.

{\it Remark.} A reader who is not a specialist of general relativity
theory, may ask the following question. Why  do we need
(something as ``fancy'' as e.g.) rotating black holes,
why is a ``simple'' Schwarzschild black hole (of sufficiently
big mass) not enough for our thought-experiment\footnote{
Cf. e.g. footnote 5 on p.83 of \cite{pit}.}?
(Instead of rotating ones, electrically charged black holes would do     
the job just as well, but this is not the issue here, since the
question is why do we need something more complex than the most
``classical'' Schwarzschild holes.) The answer is the following.

For the sake of argument, let us use Schwarzschild coordinates
for describing the spacetime outside the non-rotating black hole. Let
$\gamma_O$  and $\gamma_P$  behave as in the thought-experiment described
above (involving Kerr black holes). Now, it is true that from the 
point of view of $\gamma_P$, the clocks of $\gamma_O$ slow down so much  
that when $\gamma_O$  receives a yes (or no) answer from its computer,   
then according to $\gamma_P$'s  coordinate system, $\gamma_O$  is still
outside of the event horizon. The problem is
that to send the answer to $\gamma_O$  such that he receives it still
before hitting the singularity (which event is impossible to avoid in 
this case), $\gamma_P$ would need to use so called tachyons (FTL-signals).
Indeed, if $\gamma_P$  finishes the computation in a large enough but   
finite time, then the light $\gamma_P$ sends after $\gamma_O$ will
converge to $\gamma_O$  in a similar rate as $\gamma_O$  converges to the
singularity but will not reach $\gamma_O$  before $\gamma_O$ crosses     
it\footnote{This can be seen by looking at the
Penrose diagram of the extended Schwarzschild space-time.}.
The problem is alleviated e.g. by using rotating black holes, very   
roughly as follows. In a rotating black hole behind the event
horizon we just discussed, there is a second inner event horizon which
is a Cauchy horizon as well. If $\gamma_O$  approaches the black hole
along the orbit considered above, then not later than $\gamma_O$  reaches
the {\it second} horizon, he will meet the signal sent by $\gamma_P$. As
we indicated, making our black hole rotate is only one of the possible
solutions but this choice is strongly supported by the ``naturality''    
of rotating black holes i.e. their very possible real existence.

\section{On the physical reality of the model}
To understand if the above model is realistic from the physical point of
view, we collect properties of Malament--Hogarth space-times using
results from recent physical literature. First we summarize two important
general characteristics of Malament--Hogarth space-times. We can
state (see Lemma 4.1. and Lemma 4.3. of \cite{ear}):
\vspace{0.1in}

{\bf Proposition 5.} {\it Let $(M,g)$ be a Malament--Hogarth space-time
with a Malament--Hogarth event $p\in M$. Then $M$ is not globally
hyperbolic.

Moreover, choose any connected space-like hypersurface $S\subset M$
satisfying ${\rm im}\gamma_P\subset D^+(S)$. Then either $p\in H^+(S)$
i.e. $p$ lies on the future Cauchy horizon of $S$ or $p\notin D^+(S)$
i.e. does not belong to the future Cauchy development of $S$.}
$\Diamond$
\vspace{0.1in}

\noindent The meaning of Proposition 5 is the following. A very important
property of globally hyperbolic space-times
is that they possess a so-called initial data surface (called
Cauchy surface) i.e. fixing data of physical fields along the
Cauchy surface only (which is a three-dimensional submanifold of $M$), one
can determine the values of these fields over the whole space-time via the
corresponding field equations. The above theorem shows that a
Malament--Hogarth space-time does not possess such an initial data surface
i.e. always contains events $q\in M$ which are unpredictable even fixing
initial data on arbitrary large subsets of $M$, for example on a
space-like submanifold $S\subset M$. Especially, the Malament--Hogarth
event $p\in M$ is such an event. We met this phenomenon in the special 
case of the Kerr space-time already. The difficulties caused by this fact
will be discussed soon.

Another very important property of Malament--Hogarth space-times is
the ``infinite blue\-shift effect". Roughly speaking, as a consequence of 
the infinite time contraction seen by the observer $\gamma_O$ approaching
the Malament--Hogarth event $p\in M$, all signals
of finite energy or frequency will hit $\gamma_O$ at $p\in M$ by an
infinite amount of energy, i.e. Malament--Hogarth space-times act as
unbounded gravitational amplifiers near $p\in M$. More precisely, the
following theorem holds (Lemma 4.2. of \cite{ear}):
\vspace{0.1in}

{\bf Proposition 6.} {\it Let $(M,g)$ be a Malament--Hogarth space-time
with time-like curves $\gamma_P$ and $\gamma_O$ as in Definition 5.
Suppose that the family of null-geodesics connecting $\gamma_P$ with
$\gamma_O$ forms a two dimensional integral submanifold in $M$ in which
the order of emission from $\gamma_P$ matches the order of absorption by
$\gamma_O$. If the photon frequency $\omega_P$ is constant measured by
$\gamma_P$ (i.e. $\gamma_P$ does not stop sending signals to
$\gamma_O$) then the time-integrated photon frequency
\[\int\limits_{\gamma_O}\omega_O\dd\gamma_O=\int\limits_a^b\omega
(\gamma_O (\tau ))\sqrt{-g(\dot\gamma_O(\tau ), \dot\gamma_O(\tau
))}\:\dd\tau\]
received by $\gamma_O$ is divergent.} $\Diamond$
\vspace{0.1in}

\noindent This theorem is a trivial consequence of the assumption that the
original observer $\gamma_P$ sent an infinite amount of energy to
$\gamma_O$, since it sends signals of constant frequency $\omega_P$
throughout its infinite existence. 

Now we wish to discuss the consequences of these properties of
Malament--Hogarth space-times in the special case of the Kerr black hole
against building relativistic computers constructed in Section 3.

(1) First we are going to study the effects of the infinite
blueshift, the problem formulated in Proposition 6 above. We consider
first whether $\gamma_O$ can survive the encounter with the inner event
horizon or not. A similar but more detailed consideration like
Proposition 6 shows that near its inner horizon, the Kerr black hole
amplifies every, arbitrarily small deviation from the original vacuum
space-time structure in an unbounded amount, yielding that this horizon
rather looks like a real curvature singularity (i.e. not 
a pure ``coordinate singularity''). This phenomenon is known as the
``infinite mass-inflation'' in the physics literature and
appears if one calculates the effect of the infinitely amplified absorbed
energy on the metric near the inner horizon. At first look, in the
case of perturbations of the metric by a scalar field, this 
singularity turns out to be a scalar curvature divergence on the inner
horizon \cite{poi-isr}. This fact is usually interpreted as the
instability of the (vacuum) Kerr space-time. 
Hence, after realizing the mass-inflation phenomenon, physicists supposed
the non-traversability of the Kerr black hole. 

A more careful analysis of the situation was carried out by Ori
\cite{ori1}\cite{ori2} in the case of the Reissner--Nordstr\"om black hole
and partially in the case of the Kerr--Newman black hole, however. In
accordance with his
calculations (accepting the validity of certain technical assumptions) it
seems that despite the existence of the scalar curvature divergence,
the tidal forces remain finite moreover negligible in the case of
realistic
Kerr black holes when crossing the inner horizon. Hence although the
inner horizon (which contains the Malament--Hogarth event) is a real
curvature singularity it is only a so-called {\it weak singularity}
because the tidal forces still remain finite on it. As an
example \cite{ori2}, for a Kerr black hole of mass $M=10^7 m_\odot$ ( 
$m_\odot$ refers to solar mass) and age $T=10^6$ years (more precisely
this is the age of the initial perturbation of the Kerr black hole) the
relative distortion of an object of typical size $l$ crossing the inner
horizon is
\[{\Delta l\over l}\leq 10^{-55}.\] 
In summary, although the Malament--Hogarth event is situated in a
``dangerous'' region of the Kerr--Newman space-time, in theory at least,
it can be approached by the observer $\gamma_O$.

Next we may ask about the strong (electromagnetic) radiation absorbed 
by $\gamma_O$ during  the course of crossing the inner event horizon, as
another consequence of the blueshift effect. This problem was studied
by Burko and Ori \cite{bur-ori}. They conclude that these effects remain
also finite making it theoretically possible to survive such an encounter
for $\gamma_O$ although one may worry about the intensive pair creation
induced by the extremely high energy photons \cite{bur-ori}. Of course
these considerations require more detailed analysis in the future, see
also \cite{ori-ori}.

Summing up, we can conclude that accepting a very rough, classical
picture for the inner horizon of the Reissner--Nordstr\"om and
Kerr--Newman black holes, their traversability is reasonable. 
In our artificial computing system $G=(\gamma_P, \gamma_O)$, the
physical computer $\gamma_P$ sends a modulated light beam to $\gamma_O$.
Proposition 6 above suggests that even the (energetically) mildest answer
of $\gamma_P$ will simply destroy $\gamma_O$ by receiving an infinite
amount of energy. This is valid only if $\gamma_P$ sends electromagnetic
signals of constant
frequency through an infinite proper time (measured by $\gamma_P$), hence
Proposition 6 is not surprising. If $\gamma_P$ sends a finite signal
answering a simple ``yes'' or ``no'', the received energy by $\gamma_O$
remains finite in light of results of Burko and Ori. Hence, the
pessimistic consequences of Proposition 6 are ruled out for the
relativistic computer designed in the present paper.

(2) The stability of the circular orbit around
the Kerr black hole required for $\gamma_P$ was also studied by Kennefick
and Ori \cite{ken-ori} and Ori \cite{ori3}. They studied the effect of the
gravitational radiation of the Kerr black hole on the evolution of a point
particle moving on an initially circular orbit around the Kerr black hole.
The answer is also encouraging, the perturbation seems to be negligible
yielding the stability of stationary circular orbits, hence the computer
$\gamma_P$ can orbit around the black hole for long time (hence with
little effort for ever).

(3) We mention at this point again, that both curves $\gamma_P$ and
$\gamma_O$ are geodesics in the Kerr space-time
i.e. the acceleration along them is zero i.e. bounded. Hence in this
situation one does not have to worry about the negative consequences of a
possibly infinite acceleration, see \cite{pit}.

(4) Next we turn our attention to the consequences of Proposition 5. The
essence of this theorem is that the Malament--Hogarth event $p\in
M$ cannot be predicted even fixing initial data on the whole spatial
surface $S\subset M$ which is a Cauchy surface for the outer observer
$\gamma_P$. This fact is interpreted by Earman in \cite{ear}, p. 118 by
saying that the observer $\gamma_O$, while crossing $p\in 
M$, is able to decide whether a signal came from $\gamma_P$ or from a
possibly past singularity if and only if he is able
to perform an infinitely precise  discrimination in spatial directions
which is physically unreasonable (but although theoretically it is
allowed in classical physics). As suggested also by
Earman this problem is apparently solved by using a coding system between
$\gamma_P$ and $\gamma_O$ because in this case the information of the
result of the calculation $\gamma_P$ just completed is not carried 
simply by the direction of the light beam. But this solution is also
rejected by Earman by another ``infinitely precise discrimination''
argument (\cite{ear} p. 118) essentially based on the assumption that
$\gamma_P$ wishes to send a possibly unbounded amount of information to
$\gamma_O$. But as we clarified in the beginning of this section for our 
purposes we need to answer {\it yes or no questions} only, using a
previously fixed code. Hence the length of the message sent by $\gamma_P$
is bounded uniformly hence Earman's infinite discrimination argument is not
valid in this moment. 

But apparently, as we have seen, $\gamma_O$ must be able to perform
infinitely precise measurement of time because in our model the detection
time carries a lot of information. Notice however that this assumption is
not an extra one; it is already assumed by accepting that $\gamma_O$,
before crossing the Malament--Hogarth event $p\in M$, is always able to
detect signals from $\gamma_P$. Namely, the length of the signal received
by $\gamma_O$ tends to zero, hence very close to $p\in M$ $\gamma_O$ must
be able to detect arbitrary short signals; consequently if he can do it
certainly can measure its detection time arbitrary accurately, too;
we soon discuss how to deal with this problem (of ``infinite precision''). 

Summing up, we went through all the major possible {\it
classical} obstacles published so far against building the artificial
computing system $G=(\gamma_P, \gamma_O)$ performing
computability of the second kind, listed in \cite{ear}, \cite{pit}
and references therein. We found that these obstructions can be removed at
the classical level (hence in classical general relativity)
i.e. they do not kill the idea of designing a thought-experiment suitable
for deciding $\Sigma_1$-sets of natural numbers
(because the quoted objections do not destroy the idea of the
relativistic computer we designed in the present paper). 

As a final remark we have to emphasize again that we have omitted all
the {\it quantum effects} in our model. In general, the accuracy of time
measurement, which is required for $\gamma_O$, is not a problem in
classical physics while in quantum physics it is constrained by quantum
fluctuations. In this moment we do not
possess a satisfying theory to describe the consequences of these
quantum fluctuations in the presence of strong gravitational fields. Of
course this is because a satisfactory theory of quantum gravity has not
been formulated yet. We can do only naive considerations taking into
account the basic principles of quantum mechanics and general relativity.
Using results of Ng \cite{ng} we can say the following about the accuracy
of time measurment. Assume we have a clock with total running time
$\Delta T$ over which it can remain accurate and is capable for a time
measurment of accuracy $\Delta t$. Then one can derive an inequality
\[\Delta t\geq (\Delta T\:t^2_P)^{1/3}\]
where $t_P=\sqrt{\hbar G/c^5}\approx 10^{-43}\:sec$ is the Planck-time. In
our case we require a time measurment with an unboundedly increasing
accuracy from $\gamma_O$'s clock {\it till the Malament--Hogarth
event.} Consequently, without the violation of the above inequality, in
principle the observer $\gamma_O$ can constantly ``tune'' his clock to be
more and more accurate ($\Delta t\rightarrow 0$) for shorter and shorter
times as he approaches the Malament--Hogarth event ($\Delta T\rightarrow
0$). Possibly this clock cannot be used {\it after} the Malament--Hogarth event
but this is not a problem. But notice that interpreting the
Planck-time $t_P$ as the fundamental smallest time unit then accuracy 
beyond $t_P$ is meaningless. This might destroy the
realizability of our thought-experiment in a quantum framework. 
This means that if we use quantum gravity in place of
classical general relativity as our background theory, then we should
design our artificial computing system differently. However, since the
theory of quantum gravity does not exist yet, it seems pointless to
try to elaborate the details nowadays.

Moreover a generally accepted quantum gravitational phenomenon is the
black hole evaporation. Finally this may cause that every (Kerr) black
hole will evaporate in finite time making it impossible for $\gamma_P$ to
send signals to $\gamma_O$ in very late times. Hence these and other, yet
unknown, quantum phenomena occurring in strong gravitational fields can
eventually also annihilate our considerations.

\section{Concluding Remarks}
In this paper we have studied the physical reality of performing an
infinitude of calculations in finite time in order to answer very
interesting questions.

One of the present authors (I.N.) had discussed the various Theses
formulated in Section 2 with one of their originators L\'aszl\'o Kalm\'ar,
and he feels that Kalm\'ar would be pleased by the kind of approach taken
in the present paper.

{\bf Acknowledgement}. We are grateful to Prof. H. Kodama (Yukawa
Institute) for calling our attention to the papers of Burko and Ori and to 
Prof. H. Andr\'eka and Dr. G. S\'agi (R\'enyi Institute) for the
stimulating discussions, important remarks concerning this paper.
Thanks go to J. Madar\'asz (R\'enyi Institute) for bringing the team
together and encouraging them to write a paper. Thanks go to L. E. Szab\'o
(Department for Theoretical Physics $\&$ Department of Philosophy of
Science, E\"otv\"os University) and M. R\'edei (Department of
Philosophy of Science, E\"otv\"os University) for
keeping alive the documents of the Ames/USA course on the subjectmatter of
the paper and for continuous encouragement.

\end{document}